\documentclass[fleqn,usenatbib,useAMS]{mnras}

\usepackage{graphicx}	% Including figure files
\usepackage{amsmath}	% Advanced maths commands
\usepackage{amssymb}	% Extra maths symbols
\usepackage{multicol}    % Multi-column entries in tables
\usepackage{bm}		% Bold maths symbols, including upright Greek
\usepackage{pdflscape}	% Landscape pages
\usepackage{subcaption} % Subfigures

\usepackage[T1]{fontenc}
\usepackage{ae,aecompl}

\usepackage{newtxtext,newtxmath}
\title[Radio Galaxy Classification]{\textit{Data-Efficient Classification of Radio Galaxies}}

\author[]{Ashwin Samudre$^{1}$\thanks{Contact e-mail: \href{mailto:ashwin.samudre@gmail.com}{ashwin.samudre@gmail.com} }, Lijo T. George$^{2}$, Mahak Bansal$^{3}$ and Yogesh Wadadekar$^{2}$
\\
$^{1}$European Molecular Biology Laboratory, Meyerhofstrasse 1, 69117 Heidelberg, Germany \\
$^{2}$National Centre for Radio Astrophysics, TIFR, Post Bag 3, Ganeshkhind, Pune 411007, India \\
$^{3}$Viterbi School of Engineering, University of Southern California, 3650 McClintock Ave, Los Angeles, CA 90089, United States \\
}

\date{Accepted XXX. Received YYY; in original form ZZZ}

\pubyear{2021}

\begin{document}
\label{firstpage}
\pagerange{\pageref{firstpage}--\pageref{lastpage}}
\maketitle

\begin{abstract}
The continuum emission from radio galaxies can be generally classified into different morphological classes such as FRI, FRII, Bent, or Compact. In this paper, we explore the task of radio galaxy classification based on morphology using deep learning methods with a focus on using a small scale dataset ($\sim 2000$ samples). We apply few-shot learning techniques based on Twin Networks and transfer learning techniques using a pre-trained DenseNet model with advanced techniques like cyclical learning rate and discriminative learning to train the model rapidly. We achieve a classification accuracy of over 92\% using our best performing model with the biggest source of confusion being between Bent and FRII type galaxies. Our results show that focusing on a small but curated dataset along with the use of best practices to train the neural network can lead to good results. Automated classification techniques will be crucial for upcoming surveys with next generation radio telescopes which are expected to detect hundreds of thousands of new radio galaxies in the near future.
\end{abstract}

\begin{keywords}
galaxies: active -- techniques: image processing
\end{keywords}

%%%%%%%%%%%%%%%%%%%%%%%%%%%%%%%%%%%%%%%%%%%%%%%%%%

%%%%%%%%%%%%%%%%% BODY OF PAPER %%%%%%%%%%%%%%%%%%

\section{Introduction}

% ABOUT RADIO GALAXIES
The radio continuum sky contains a wide variety of sources. Some are compact sources while others are extended. Compact sources are unresolved sources with a single source component whereas extended sources are those whose sizes are greater than the effective resolution of the telescope.

\begin{figure}
    \centering
    \includegraphics[width=0.5\linewidth]{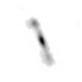}\\
    \includegraphics[width=0.5\linewidth]{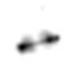}
    \caption{A typical example of an FRI (top) and FRII (bottom) type radio galaxy from our sample. Each image is $270\times270$ arcsec on a side.}
    \label{fig:FR-sample}
\end{figure}

% RADIO GALAXY MORPHOLOGY CLASSES
In some galaxies, emission from the central Active Galactic Nucleus (AGN) is seen in the form of relativistic jets which manifest as radio synchrotron emission. These radio jets usually come in pairs and extend outwards in diametrically opposite directions. When they collide with the denser parts of the intergalactic medium, they give rise to the radio lobes. The morphology of the twin jets and radio lobes is used to broadly classify radio galaxies into FRI and FRII types \citep{fanriley74}. In FRI galaxies, the peak of the emission is closer to the central core and the emission brightness decreases farther away from the centre. On the other hand, the radio emission in FRII galaxies increases in brightness outwards culminating in bright hotspots at the outer edge of the jet emission. See Figure~\ref{fig:FR-sample} for typical examples of FRI and FRII galaxies drawn from our sample.
 
Apart from these two broad classes, some radio galaxies also show a more complex morphology. If the angle between the jets in the radio galaxy is significantly less than $180^{\circ}$, then it can be classified as a Bent-Tail (BT) galaxy. BT galaxies can be further subclassified as Narrow-Angle Tail (NAT) or Wide Angle Tail (WAT) galaxies based on the angle between the two jets \citep{rudwen76}. In NATs, the jets are so bent that the entire radio structure lies on one side of the optical galaxy \citep{miley72} whereas in WATs the angle of projection between the jets is $\sim 60^{\circ}$ \citep{dehghan14}. Bent Tail galaxies are believed to form in dense environments such as galaxy clusters. The bent tail phenomenon is usually attributed to the large ram-pressure exerted on the jets as a result of the motion of the host galaxy through the cluster medium \citep{miley72,rudwen76,burns90} or due to the turbulent movement of the ICM (Intra-Cluster Medium) during a merger \citep{burns90,burns96,roettiger96}. As such, they can be effectively used as tracers for the presence of galaxy clusters, particularly at high redshifts \citep{blanton00,blanton03,mao09,mao10,wington11,dehghan14}. It is expected that the upcoming Evolutionary Map of the Universe (EMU) survey \citep{norris11} will be able to detect many new galaxy clusters via this method.

% CURRENT CATALOGS & FUTURE DETECTIONS
Traditionally, radio galaxy classification has been carried out by visual inspection of radio continuum images. Large area radio surveys like the NVSS \citep{nvss98}, FIRST \citep{first95}, SUMSS \citep{sumss99} have been used to identify and classify radio galaxies based on their morphology. The Radio Galaxy Zoo (RGZ) project aims to use citizen scientists to assist in the identification and classification of candidate radio galaxies \citep{rgz15}. In the future, next generation radio telescopes like MeerKAT \citep{2016mks..confE...1J}, ASKAP \citep{askap08} and eventually the SKA \citep{ska04} with significantly increased sensitivity will detect millions of new radio galaxies. Given these astronomical numbers, manual classification of radio galaxies would be highly impractical, if not impossible. As a result, astronomers need to develop an automated way to precisely classify radio galaxies based on their morphology.

% EXISTING USE OF ML and DL in ASTRONOMY
Automated classification algorithms using machine learning and deep learning techniques are extremely useful for this purpose. Machine learning techniques have been widely applied to astronomical problems in recent years. See \citet{Fluke2020} for a recent review.

Classification of optical galaxies using artificial neural networks has been a particularly active field of research \citep{philip02,dieleman15,hc15,ostrovski17,schawinski17,hocking18,martin20,cheng20}. \cite{aniyan17} were the first to employ a CNN (Convolutional Neural Network) to classify radio galaxies using FIRST survey images. Since then several other researchers  \citep{wathela18, 2018MNRAS.476..246L, ma19, 10.1093/mnras/stz1289, wu19, tang19} have used different variations of the standard CNN architecture to attempt radio galaxy classification.  

%Deep learning primer
Recently, deep learning \citep{lecun2015deeplearning} algorithms have shown remarkable efficacy in image classification and speech recognition tasks across diverse fields.  McCulloch \& Pitts introduced the McCulloch-Pitts Neuron \citep{mcculloch43a} in 1943, which is considered as the initial version of the Artificial Neural Network. Further work by \cite{hebb-organization-of-behavior-1949}, \cite{rosenblatt1958perceptron} laid the foundation for modern neural networks. The current deep learning era started with the introduction of Deep Belief Networks by \cite{hinton06} (see \citet{DBLP:journals/corr/WangRX17} for a comprehensive history of deep learning). The term deep learning itself is associated with the collection of new techniques mainly based on Neural Networks. In these networks, inputs are fed into the input layer propagating through one or more hidden layers and finally linking to the output layer. A layer is mainly a set of nodes, sometimes referred to as units, connected to the immediately preceding and following layers. The input layer nodes are associated with the inputs from the dataset of interest. In deep neural networks, multiple layers are present (hence the term deep) and starting with the second layer, each layer essentially performs the feature construction based on the output of the previous layers which in turn acts as the input to the current layer. The output from the final layer is compared with the expected output and the difference (also called the `error') is propagated backward (called backpropagation \citep{Rumelhart:1986we}), through all the layers to adjust the units in each layer and the process is repeated. This backpropagation process helps in better optimisation of the network for each input dataset.

In this work, we have implemented a method to perform automated morphological classification of radio galaxies using deep learning into four important classes - FRI, FRII, Bent, and Compact. The novel feature of this work is the focus on creating a classifier with fewer data samples while still striving for maximum accuracy. \cite{roh2018survey} provide an overview of the role of data in machine learning and deep learning techniques. An approximate rule of thumb is that we need at least 1000 samples per class in the task. The State of the Art results for Computer Vision tasks \citep{DBLP:journals/corr/abs-1905-00546, 2019arXiv191211370K} based on ImageNet \citep{deng2009imagenet} and MNIST \citep{lecun2010mnist} datasets have used architectures like ResNet \citep{he2016deep}, DenseNet \citep{huang2017densely} for  ideal performance but the sample size for these tasks has been at least 1000 images per class. \cite{8599448} discuss the impact of data sample size in image classification tasks and show that sample size affects the performance as we increase the number of classes in the task. For the transfer learning approach on image classification tasks, \cite{10.1007/978-3-319-46349-0_5} discuss that in case of a limited data sample, training only specific layers can provide performance similar to a model trained from scratch with a large dataset. 

The previously mentioned approaches to build a classifier for radio galaxies have used a large amount of data (at least 10000 samples in each class) for experiments. Where necessary, sample sizes have been increased with the use of techniques like oversampling and data augmentation \citep{NIPS2012_4824}. In \citet{wathela18}, the authors achieved an overall accuracy above 94\% (with the same four classes as in this work - FRI, FRII, Bent, Compact used for classification) using more than 21000 samples. \citet{aniyan17} performed a 3 class (FRI, FRII and Bent) classification using a small original dataset (716 total samples) but above 25000 samples for experiments with substantial use of augmentation. \citet{ma19} implemented an unsupervised pre-training technique on the unlabeled samples and later use this pre-trained model to classify the labeled samples as a 6 class classification task. They use 1442 labeled and 14245 unlabeled samples and at least 4900 samples per class in case of experiments with augmentation in their two-step approach. \citet{2018MNRAS.476..246L} performed 4 class (Compact, Single extended, Two-Component extended and Multi-component extended) classification using a total of 141032 samples via augmentation (21933 original samples) for 4 classes. \citet{wu19} used object detection and classification techniques while using more than 10000 samples for their experiments. They also leveraged the transfer learning technique by loading pre-trained model weights based on ImageNet dataset. \citet{tang19} implemented transfer learning approach while maintaining the idea of domain transfer. They used NVSS \citep{nvss98} and FIRST \citep{first95} catalog samples and performed 1) pre-training of the model on NVSS data and transferred the weights for FIRST data and 2) pre-trained model on FIRST data, transferred weights for NVSS data. Their task is a 2 class (FRI and FRII) i.e. binary classification task. \citet{10.1093/mnras/stz1289} performed a 3 class (Unresolved, FRI and FRII) classification using a model pre-trained on ImageNet dataset and 15936 total image samples.

We have limited our data sample size to $\sim$2000 images for initial experiments and $\sim$2700 images for the experiments with the balanced classes. As we increase the number of data samples, we can capture the distribution of the data in a better way. However, obtaining a large number of samples is not always feasible. Thus, we need to develop algorithms that can perform better even with a limited amount of data. In this work, we apply few-shot learning and transfer learning techniques. Our results show that with the use of a carefully curated dataset and a well-designed neural network architecture, higher accuracy can be obtained even with less amount of data for this particular classification task.

The paper is organised as follows: Section~\ref{sec:methodology} presents the overall approach towards the problem and includes the data sample construction used for the experiments, the initial approach of few-shot learning technique and its intuition to work efficiently with less amount of data, transfer learning introduction with a description of pre-trained models and the specific architecture used for our experiments. Section~\ref{sec:expresults} presents the experimental details where we try out and apply the optimum parameters to obtain the results. In Section~\ref{sec:discussion}, we discuss the results, choice of hyperparameters and the scope for further work. Finally, in Section~\ref{sec:conclusion}, we provide a self-contained conclusion listing the main results of the paper.

\section{Methodology}
\label{sec:methodology}
\subsection{Data Sample construction}
\label{sec:data}

Our radio galaxy sample was compiled using data from several catalogs. We used the catalog by \cite{proctor11} for bent-tail galaxies (BENT). For compact sources (Compact) we combined the FR0 catalog by \cite{baldi17} and the CoNFIG (Combined NVSS-FIRST Galaxies) sample \citep{gendre08, gendre10}. We only used the sources classified as ``Compact (C)'' for this sample. The sample of FRI galaxies included the FRICAT by \cite{capetti17a} as well as the sources classified as FRI in the CoNFIG sample. Similarly, the FRII sample of galaxies was combined from the FRIICAT by \cite{capetti17b} and the sources classified as FRII in the CoNFIG sample. We did not use sources classified as ``unresolved'' from the CoNFIG catalog. We constructed image cutouts of size  $270\times270$ arcsec for all galaxies in our sample. Each image was converted from the original FITS format to the JPEG format using {Astropy}\footnote{\url{https://www.astropy.org}}.

The complete data sample initially contained 412 Bent, 332 Compact, 321 FRI, and 606 FRII sources. We manually inspected the VLA FIRST survey image for every galaxy in our sample and discarded a few galaxies based on 1) Presence of multiple sources in a single image, 2) High noise component in the image, 3) Source present but not centered in the image cutout. Through this manual procedure, we removed 7 samples from the Bent set, 26 samples from the Compact set, 32 samples from the FRI set and 29 samples from the FRII set.
About $\sim$25\% samples from each class were used for validation purposes. The test set images were obtained separately per category. The numbers of objects in each category are listed in Table~\ref{table:Table 1}.

Analogously to the approach used by \cite{aniyan17}, some preprocessing was performed on the images before being used. Noise statistics were calculated in every image and all pixels below $3\sigma$ were clipped. Each image was rescaled to a size of $224 \times 224$ pixels to make it compatible with ImageNet stats.

Our data sample contains data imbalance between the classes. This leads to an uneven representation of the different classes during the learning process of the model and the results tend to be biased towards the majority class \citep{5128907}. The methods suggested in the literature for handling the imbalance include majority undersampling, minority oversampling, and SMOTE like techniques (\cite{Krawczyk2016LearningFI}, \cite{5949434}, \cite{chawla2002smote}, \cite{10.1007/11538059_91}). In \cite{branco2015survey}, authors analyze the different methods devised to handle the imbalance of classes in data, which primarily suggest balancing the dataset or algorithm level improvement. Since we are using a small scale dataset, we opt for upsampling the minority classes instead of the downsampling approach. This fits our purpose of removing the class imbalance, while still avoiding the large increase in the data sample size like was done by other researchers for the radio galaxy classification task.

We handle the class imbalance relative to the largest class (FRII) via upsampling the other 3 classes. The upsampling is performed via duplication of randomly selected samples from the original set and combining these with the original set. The numbers of objects in the balanced dataset for each category are listed in Table~\ref{table:Table 2}.

With this data sample, we initially implemented the Few-Shot learning approach and used Twin Networks for classification purposes.

\begin{table}
    \centering
    \caption[Centre]{Final data sample distribution: number of total images, training, validation and testing images (Original set).}
    \label{table:Table 1}
    \begin{tabular}{lcccr} % four columns, alignment for each
        \hline
        Type & Number of samples & Train & Validation & Test\\
        \hline
        Bent & 508 & 305 & 100 & 103\\
        Compact & 406 & 226 & 80 & 100\\
        FRI & 389 & 215 & 74 & 100\\
        FRII & 679 & 434 & 144 & 101\\
        \hline
        Total & 1982 & 1180 & 398 & 404\\
        \hline
    \end{tabular}
\end{table}

\begin{table}
    \centering
    \caption[Centre]{Final data sample distribution with balanced classes: number of total images, training, validation and testing images.}
    \label{table:Table 2}
    \begin{tabular}{lcccr} % four columns, alignment for each
        \hline
        Type & Number of samples & Train & Validation & Test\\
        \hline
        Bent & 680 & 433 & 144 & 103\\
        Compact & 675 & 431 & 144 & 100\\
        FRI & 674 & 430 & 144 & 100\\
        FRII & 679 & 434 & 144 & 101\\
        \hline
        Total & 2708 & 1728 & 576 & 404\\
        \hline
    \end{tabular}
\end{table}

\subsection{Performance metrics}

Although accuracy is often the primary evaluation metric for the performance of classification problems, to quantify the performance of our network\textquotesingle s classification ability, we use other metrics such as Precision, Recall, and F1-score. These metrics provide detailed insights based on the performance of the classifier for each class.  The precision P (purity) and recall R (completeness) are defined as follows:

\begin{equation} \label{precision}
    \begin{gathered}
        Precision (P) = \frac{TP}{TP + FP}
    \end{gathered}
\end{equation}

\begin{equation} \label{recall}
\begin{gathered}
Recall (R) = \frac{TP}{TP + FN}
\end{gathered}
\end{equation}
\newline
The F1 score is interpreted as the weighted average of precision and recall. The accuracy is the overall accurate prediction by the classifier across all the classes.

\begin{equation} \label{F1-score}
\begin{gathered}
F1-score = \frac{2 \times \text{Precision} \times \text{Recall}}{\text{Precision} + \text{Recall}}
\end{gathered}
\end{equation}

\begin{equation} \label{accuracy}
\begin{gathered}
Accuracy = \frac{TP + TN}{TP + FP + TN + FN}
\end{gathered}
\end{equation}
\newline
where TP=true positive, TN=true negative, FP=false
positive, FN=false negative.
\begin{description}
\item[$\bullet$] True Positive (TP) is when the source is predicted as `x' and it is actually `x'.
\item[$\bullet$] False Positive (FP) is when the source is predicted as `x' and it is not actually `x'.
\item[$\bullet$] True Negative (TN) is when the source is predicted as not `x' and it is actually not `x'.
\item[$\bullet$] False Negative (FN) is when the source is predicted as not `x' and it is actually `x'.
\end{description}
The precision, recall, F1 score, and accuracy metrics were calculated for both the validation and test data sets to assess the performance in each experiment with the corresponding model.

\subsection{Few-Shot Learning}

A variety of algorithms in deep learning have performed efficiently in terms of pattern recognition for image data. These algorithms often need large amounts of data and break down when forced to make predictions about data for which little supervised information is available. The ability to acquire and recognize new patterns is strong in humans. In particular, people seem to be able to understand new concepts quickly and then recognize variations on these concepts in future cases \citep{Lake_oneshot}. The deep learning algorithms are unable to learn the features for a class from a single image and generalize this information for the unseen images from the same class. Few shot learning is a practice used to provide the machine with a quick understanding of the different images and their corresponding classes using limited amount of data.

Few shot learning techniques are devised to classify image samples correctly even when the availability of supervised data is limited. It is a procedure that resembles human-like learning. 

There are multiple approaches to implement and test the performance of Few shot learning amongst which we choose a {Twin Network}\footnote{Twin networks are referred to as Siamese networks in the literature. We feel that such terminology is inappropriate, and should not be used.} architecture with a supervised learning approach.

\subsubsection{Advantages of Few Shot Learning (FSL):}

Few Shot Learning can help in the reduction of the data-gathering efforts as well as the computation time and cost that arises with large amounts of data. The image classification accuracy ({\it P}) improves with the experience ({\it E}) obtained by a few labeled images for each class of the target ({\it T}), and the prior knowledge extracted from other classes, such as raw images to co-training \citep{colt98/blum}. Methods that succeed in this task usually have higher generality; therefore they can be easily applied for tasks with large training samples.

Few Shot Learning can help in training suitable models for rare cases when the amount of correctly labeled data are limited. Although this is a common situation in many astronomical situations, to our knowledge, this work represents the first application of few shot learning in astronomy.

\subsubsection{Twin networks}

\begin{figure}
\centering
%\begin{subfigure}{.5\textwidth}
% \centering
 \includegraphics[width=1.0\linewidth]{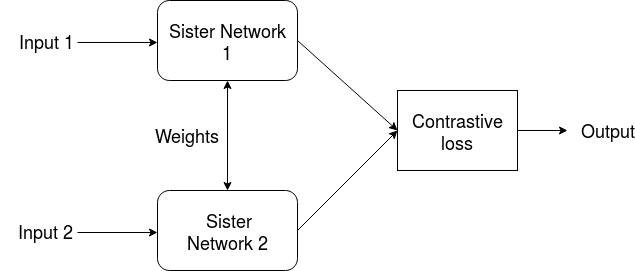}
 \caption{Overview of a simple Twin Network.}
 \label{fig:Siamese_arch}
%\end{subfigure}%
\end{figure}

Twin networks \citep{koch2015siamese} are neural networks containing two or more identical subnetwork components. A Twin network can be visualized as shown in Figure~\ref{fig:Siamese_arch}. The network is called ‘Twin’ as the subnetworks have identical architecture and the weights are also shared among them. Twin network architecture is based on the idea that it can learn useful data descriptors that assist to compare the inputs of the respective subnetworks. The network handles different types of input such as numerical data (in this case the subnetworks are usually formed by fully-connected layers), image data (with Convolutional Neural Networks as subnetworks, as in the current case) or even sequential data such as sentences or time signals (with Recurrent Neural Networks as subnetworks). \newline

\begin{table}
 		\centering
 			\begin{tabular}{lccr}
 				\hline
 				Num & Layer type & Output Shape & Parameters\\
 				\hline
 				0 & Input & (1, 224, 224) & 0 \\
 				\hline
 				1 & Convolutional layer & (64, 224, 224) & 640 \\
 				  & ReLU & (64, 224, 224) & 0 \\
 				\hline
 				2 & BatchNormalization layer & (64, 224, 224) & 128 \\
 				\hline
 				3 & Convolutional layer & (32, 224, 224) & 18464 \\
 				  & ReLU & (32, 224, 224) & 0 \\
 				\hline
 				4 & BatchNormalization layer & (32, 224, 224) & 64 \\
 				\hline
 				5 & Convolutional layer & (16, 224, 224) & 4624 \\
 				  & ReLU & (16, 224, 224) & 0 \\
 				\hline
 				6 & BatchNormalization layer & (16, 224, 224) & 32 \\
 				\hline
 				7 & Convolutional layer & (8, 224, 224) & 1160 \\
 				  & ReLU & (8, 224, 224) & 0 \\
 				\hline
 				8 & BatchNormalization layer & (8, 224, 224) & 16 \\
 				\hline
 				9 & Convolutional layer & (8, 224, 224) & 584 \\
 				  & ReLU & (8, 224, 224) & 0 \\
 				\hline 
 				10 & BatchNormalization layer & (8, 224, 224) & 16 \\
 				\hline
 				11 & Linear & (500) & 10,000,500 \\
 				   & ReLU & (500) & 0 \\
 				\hline
 				12 & Linear & (500) & 250,500 \\
 				   & ReLU & (500) & 0 \\
 				\hline
 				13 & Linear & (4) & 2004 \\
 			\end{tabular}
 			\caption{Twin Network architecture used in our experiments.}
 	\label{table:Siamese-Net}
 	\end{table}

\noindent\textbf{Model Architecture}
\smallskip

We used five convolution layers with batch normalization and two fully connected layers with ReLU activation function as the architecture for the subnetworks of the Twin Network as described in Table~\ref{table:Siamese-Net}. The loss function used here is Contrastive Loss Function \citep{Hadsell:2006:DRL:1153171.1153654}. In training deep neural networks, an activation function is used to define the output of a given node. It helps to avoid the linearity between the outputs of subsequent layers and thus the network can learn complex non-linear patterns from the data. Batch normalization \citep{DBLP:journals/corr/IoffeS15} (also referred to as BatchNorm) is another technique to normalize the inputs to each layer during training, which is performed using mean and standard deviation of the values in the current batch. This helps to reduce the difference in input parameters (obtained from a variety of input samples) to each layer. A traditional CNN consists of two parts: a feature learning section consisting of convolutional layers and a classification section in the end consisting of fully connected layers. A subnetwork of twin networks only contains the former to learn embeddings of the image provided as input. The goal of the Contrastive Loss Function (see equation~\ref{ContrastiveLoss}) is to receive the embeddings from the two subnetworks and output a lower distance if the images belong to the same class and higher distance if they belong to different classes. \newline

In Equation~\ref{ContrastiveLoss} for contrastive loss, sim$(z_i, z_j)$ is the similarity function (cosine distance between the vectors) and $\tau$ is the temperature regularization factor.

\begin{equation} \label{ContrastiveLoss}
    l_{i,j} = - \log \frac{\exp(\text{sim}(\boldsymbol{z_i,z_j})/\tau)}{\sum^{2N}_{k=1~[k\neq i]} \exp(\text{sim}(\boldsymbol{z_i,z_k})/\tau)}
\end{equation}

\smallskip\smallskip

\noindent\textbf{Training process}
\smallskip

The training process for the Twin network involves providing two input images (one for each subnetwork) and a label to the network. The label defines if the images belong to the same class (positive) or different classes (negative). The embeddings from the subnetworks are passed to the Contrastive Loss Function along with the label (i.e positive or negative) to learn and differentiate between pairs of images belonging to the same class and different classes. Ideally, the trained Twin subnetworks output similar embeddings for images of the same class and different embeddings for images of different classes. \newline

\noindent\textbf{Testing process}
\smallskip

Unlike traditional CNNs where the trained model can classify the image passed to it, trained subnetworks of the Twin Network model can only output embeddings of the images passed to them. Therefore we use the k-nearest neighbors (KNN) technique to classify images after passing them through a trained Twin network. The test image is compared to the embeddings of all the images in training data (stored earlier) and the class of images that are closest to this image (k-nearest neighbor) is assigned to the test image.

\subsubsection{\textbf{Few-Shot Learning experiments and results}}

We used the open source PyTorch \citep{paszke2019pytorch} framework, Python version 3.6.7 for our experiments and Nvidia P100 graphics processing unit (GPU) for the hardware support.

We implemented a customized data loader in PyTorch to feed input pairs to the Twin Networks.

Twin Networks perform pairwise learning and learn to find the similarity between the pair of images as opposed to a traditional CNN that learns to classify a single image predominantly via loss function like cross-entropy loss.

We perform the experiments with two different samples: the original imbalanced dataset and the balanced dataset obtained via upsampling of minority classes with respect to the majority class. Since we will be referring to the two sets of experiments frequently, we use the following notations hereafter. Version 1: original imbalanced dataset experiments, Version 2: balanced dataset experiments.

The performance of Twin Networks on the original dataset provided an overall accuracy of 71.1 $\pm$ 0.4. The performance metrics are listed in Table~\ref{table:Test_siamese_original}. We refer to mean over all classes as `avg/ total' in the validation and test metrics tables hereafter.

Further, we tried the same network on a balanced dataset. The overall accuracy improved to a value of 73.9 $\pm$ 0.5. Performance metrics are listed in Table~\ref{table:Test_siamese_balanced}.

\begin{table}
  
    \centering
    \caption[Centre]{Test metrics: Twin Networks, original dataset}
    \label{table:Test_siamese_original}
    \begin{tabular}{lcccr} % four columns, alignment for each
        \hline
        Class & Precision & Recall & F1-Score & Support\\
        \hline
        Bent & 53.2 $\pm$ 0.3 & 45.1 $\pm$ 0.5 & 48.9 $\pm$ 0.3 & 103\\
        Compact & 96.3 $\pm$ 0.5 & 89.1 $\pm$ 0.8 & 92.9 $\pm$ 0.3 & 100\\
        FRI & 66.6 $\pm$ 0.6 & 83.1 $\pm$ 0.9 & 73.8 $\pm$ 0.8 & 100\\
        FRII & 68.4 $\pm$ 1.3 & 66.8 $\pm$ 0.5 & 67.7 $\pm$ 0.7 & 101\\
        \hline
        avg/ total & 70.6 $\pm$ 1.1 & 71.2 $\pm$ 0.7 & 70.8 $\pm$ 0.6 & 404\\
        \hline
    \end{tabular}
\end{table}

\begin{table}
  
    \centering
    \caption[Centre]{Test metrics: Twin Networks, balanced dataset}
    \label{table:Test_siamese_balanced}
    \begin{tabular}{lcccr} % four columns, alignment for each
        \hline
        Class & Precision & Recall & F1-Score & Support\\
        \hline
        Bent & 61 $\pm$ 0.3 & 45.0 $\pm$ 0.7 & 51.5 $\pm$ 0.5 & 103\\
        Compact & 97.9 $\pm$ 0.8 & 88.4 $\pm$ 0.6 & 93 $\pm$ 0.4 & 100\\
        FRI & 66.2 $\pm$ 0.4 & 88.5 $\pm$ 0.5 & 75.6 $\pm$ 0.3 & 100\\
        FRII & 72.4 $\pm$ 0.4 & 73.9 $\pm$ 0.2 & 73.2 $\pm$ 0.3 & 101\\
        \hline
        avg/ total & 74.2 $\pm$ 0.6 & 74.0 $\pm$ 0.5 & 74.1 $\pm$ 0.4 & 404\\
        \hline
    \end{tabular}
\end{table}

Although the results of Twin Networks implementing Few Shot Learning are encouraging, they are not as good as the state-of-the-art techniques described in recent literature. We wanted to see if the use of other more sophisticated techniques produces better results than the simple architecture of the Twin network. We therefore switched to using the Transfer Learning approach using the same input data.

\begin{table*}
    \centering
    \caption[Centre]{The DenseNet-161 architecture}
    \label{table:densenet-161}
    \begin{tabular}{lccccccr} % five columns, alignment for each
        \hline
        \# Layer & Function & Input tensor size & Activation & Stride & Output tensor size & No. of parameters & \# Block \\
        \hline
        %input
        0 & Input & 3 x 224 x 224 & - & - & - & 0 & \\
        
        % Conv1 layer
        1 & \textbf{conv} & 3 x 224 x 224 & ReLU & 2 & 96 x 112 x 112 & 14112 & - \\
        1 & pooling & 96 x 112 x 112 & - & 2 & 96 x 56 x 56 & 0 & \\
        \hline
        % Dense block 1
        2-13 & [1 x 1 \textbf{conv}, 3 x 3 \textbf{conv}] x 6 & 96 x 56 x 56 & ReLU & 2 & 48 x 56 x 56 & 82944 & 1 \\
        \hline
        % Transition layer 1
        14 & \textbf{conv} & 384 x 56 x 56 & ReLU & 2 & 192 x 56 x 56 & 73728 & - \\
        14 & pooling & 192 x 56 x 56 & - & 2 & 192 x 28 x 28 & 0 & \\     
        \hline
        % Dense block 2
        15-38 & [1 x 1 \textbf{conv}, 3 x 3 \textbf{conv}] x 12 & 192 x 28 x 28 & ReLU & 2 & 48 x 28 x 28 & 82944 & 2 \\
        \hline
        % Transition layer 2
        39 & \textbf{conv} & 768 x 28 x 28 & ReLU & 2 & 384 x 28 x 28 & 294912 & - \\
        39 & pooling & 384 x 28 x 28 & - & 2 & 384 x 14 x 14 & 0 & \\
        \hline
        % Dense block 3
        40-111 & [1 x 1 \textbf{conv}, 3 x 3 \textbf{conv}] x 36 & 384 x 14 x 14 & ReLU & 2 & 48 x 14 x 14 & 82944 & 3 \\        
        \hline
        % Transition layer 3
        112 & \textbf{conv} & 2112 x 14 x 14 & ReLU & 2 & 1056 x 14 x 14 & 2230272 & - \\
        112 & pooling & 1056 x 14 x 14 & - & 2 & 1056 x 7 x 7 & 0 & \\
        \hline
        % Dense block 4
        113-160 & [1 x 1 \textbf{conv}, 3 x 3 \textbf{conv}] x 24 & 1056 x 7 x 7 & ReLU & 2 & 48 x 7 x 7 & 82944 & 4 \\
        \hline
        % Classification layer
        161 & BatchNorm2d & 48 x 7 x 7 & - & - & 2208 x 7 x 7 & 4416 & - \\
        161 & avgpool & 2208 x 7 x 7 & - & 2 & 2208 x 1 x 1 & 0 & - \\
        161 & maxpool & 2208 x 1 x 1 & - & 2 & 2208 x 1 x 1 & 0 & - \\
        161 & Flatten & 2208 x 1 x 1 & - & - & 4416 & 0 & - \\
        161 & BatchNorm1d & 4416 & - & - & 4416 & 8832 & - \\
        161 & Dropout & 4416 & - & - & 4416 & 0 & - \\
        161 & Linear & 4416 & ReLU & - & 512 & 2261504 & - \\
        161 & BatchNorm1d & 512 & - & - & 512 & 1024 & - \\
        161 & Dropout & 512 & - & - & 512 & 0 & - \\
        161 & Linear & 512 & - & - & 4 & 2052 & - \\

        \hline
    \end{tabular}
\end{table*}

\subsection{Transfer Learning}

Transfer learning \citep{pratt1993discriminability} is a technique in machine learning which focuses on storing the knowledge gained while solving one problem and applying it to a different but related problem (See \citet{2018arXiv180801974T} for an overview of deep transfer learning). The definition of transfer learning is given in terms of domain and task. The domain D consists of: a feature space X and a marginal probability distribution $P(X)$, where $X = {x_1,...,x_n}$ belong to $X$. Given a specific domain, $D = \{X, P(X)\}$, a task $T$ consists of two components: a label space $Y$ and an objective predictive function $f$ (denoted by $T = \{Y, f\}$), which is learned from the training data consisting of pairs $\{x_i, y_i\}$, where $x_i$ belongs to $X$ and $y_i$ belongs to $Y$. The function $f$ can be used to predict the corresponding label, $f(x)$, of a new instance $x$ \citep{pan2009survey}. In retrospect, transfer learning is the ability to utilize existing knowledge from the source learner in the target task.\\

\noindent\textbf{Pre-trained models}
\smallskip

One important requirement for transfer learning is the availability of models that perform well on source tasks. Pre-trained models are trained on a large amount of data with a variety of classes. The weights of such a network are set to obtain the low-level features like shapes, edges identified by the initial layers of the network efficiently. In this work, we use the method of transfer learning by loading weights pre-trained on the ImageNet dataset using the DenseNet architecture and its variants. We also experimented with some other architectures than the architecture presented here with no significant improvement in the performance. However, we do not claim to have explored the complete parameter space.

We also note that we could not get satisfactory results with a custom convolutional neural network architecture and thus opted for the DenseNet architecture.

\subsubsection{DenseNet Architecture}
In Convolutional Neural Networks, the input image goes through multiple convolutions to obtain high-level features. Feature maps are the collective output of the convolution operation applied over the whole input image by all the convolution filters. Skip connections are the connections to the current layer from layers other than the immediately preceding layer. In DenseNet, the feature maps are concatenated from each of the previous layers along with the addition of the skip connections obtained from the previous layers. Thus, each layer receives collective knowledge from all the previous layers. Traditional CNN with `L' layers have `L' connections between the subsequent layers whereas DenseNet has $L\times(L+1)/2$ direct connections. This helps to avoid the learning of redundant feature maps and leads to a reduction in the requirement of total parameters of the network. The important element in DenseNet is the dense block which is composed of the Batch Normalization layer \citep{pmlr-v37-ioffe15}, ReLU activation function \citep{nair2010rectified} and $3\times3$ convolution layer and the dimensions of the feature maps remain constant within a block. The dense blocks contain transition layers in between them consisting of Batch Normalization, Convolution, and Pooling layers. Growth rate (K) of the network due to concatenation of the feature maps at each layer `L' can be formulated as $K(L) = K(0) + K\times(L-1)$ and it regulates the amount of information added to the network at each layer. At the end of the last block, global average pooling is performed and then a softmax activation is attached to get the final result.

\smallskip
\noindent The DenseNet architecture has two important advantages:
\begin{enumerate}
    \item The availability of connections between all the layers helps to propagate the error or the gradient more directly. The earlier layers can get feedback from the final layer rapidly.
    \item The input received by the current layer has knowledge from all the preceding layers leading to diversified features and richer patterns. This helps to obtain smoother decision boundaries. It also explains why DenseNet performs well when training data are insufficient - the central focus for our experiments.
\end{enumerate}

\subsubsection{\textbf{Experimental setup and parameter selection}}
\label{parameters}
We used the Fastai \citep{Howard_2020} library, Python version 3.6.7 for our experiments and Nvidia P100 graphics processing unit (GPU) for the hardware support. We set the batch size to 32 and the image size to $224\times224$, the data samples are normalised using the ImageNet stats.

In order to select the optimal hyperparameters such as the optimizer and learning rate scheduler, we experimented with three optimisers - Stochastic Gradient Descent (SGD), Adam and AdamW - along with three learning rate schedulers - Cosine Annealing scheduler, Multistep scheduler and Cyclical scheduler. Based on the maximum accuracy reached (see  Table~\ref{table:Hyperparam-setup}), we selected AdamW optimizer with Cyclical learning rate scheduler in further experiments.

\begin{table}
  
    \centering
    \caption[Centre]{Hyperparameters setup: Maximum accuracy on validation set with the selected optimizer and LR scheduler}
    \label{table:Hyperparam-setup}
    \begin{tabular}{lccr} % four columns, alignment for each
        \hline
         & CosAnn & Multistep & Cyclical \\
        \hline
        Adam & 79.5 $\pm$ 0.4 & 85.5 $\pm$ 0.1 & 91.4 $\pm$ 0.3 \\
        SGD & 84.7 $\pm$ 0.3 & 84.9 $\pm$ 0.5 & 90`.9 $\pm$ 0.4 \\
        AdamW & 88.2 $\pm$ 0.4 & 89.4 $\pm$ 0.3 & 91.8 $\pm$ 0.3 \\
        \hline
    \end{tabular}
\end{table}

We use the learning rate finder method based on the cyclical learning rates \citep{2015arXiv150601186S} to obtain the optimal learning rate. This method performs a mock training run starting with a smaller value for the learning rate which is modified after each batch. It searches the maximum learning rate at which there is still some improvement in terms of a decrease in the error. The learning rate either increases or decreases based on the outcome from the latest batch in the epoch. Large neural networks take a considerable amount of time for the training and the choice of the learning rate substantially affects this process. Using cyclical learning rates technique helps in the selection of a greater value for learning rate while still maintaining the improvement of the model. Based on the data distribution, after using the learning rate finder method, we set the learning rate for each experiment individually. We use a weight decay of $1e-3$ and categorical cross-entropy (equation~\ref{CELoss}) as the cost function for all the tasks. In equation~\ref{CELoss}, $x,y$ denote the classes and observations respectively, $t_{x,y}$ represent targets and $p_{x,y}$ represent the predictions. We have limited the training to 10 epochs for our experiments.
\begin{equation} \label{CELoss}
    \begin{gathered}
        L_x = - \sum_{y}t_{x,y} log(p_{x,y})    
    \end{gathered}
\end{equation}

The growth rate {\it k}  in the DenseNet is the number of output feature maps of a layer and plays an important role in propagating the information through the network. We tried different versions of the DenseNet architecture and the best performance for our task was shown by the DenseNet-161 version. Hereafter, we focus on fine-tuning the selected version. The number 161 refers to the total number of layers in the particular version of the architecture. This can be described as follows: DenseNet-161 consists of 4 dense blocks and hence 3 transition layers between the blocks. There is one convolution layer at the beginning and one classification layer (fully connected layer) at the end. The four dense blocks consist of 12, 24, 72, 48 `conv' layers respectively where `conv' layer is combined BatchNorm, ReLU activation and Convolution. Table \ref{table:densenet-161} describes the functions and the classification head at the end for the model. While the total number of parameters in DenseNet-161 is 28,745,412 freezing initial layers and training only the last layer reduces the number of trainable parameters to 2,493,348.

Initially, we train the last layer for 1-2 epochs. The intuition behind training this way is to train only the fully connected layers completely and freezing the other layers for faster training. This way, we use the weights obtained from the pre-trained model and check the performance of the model. Gradually unfreezing the layers works better than unfreezing and retraining all layers of the network \citep{yosinski2014transferable, DBLP:journals/corr/abs-1801-06146}.
Here, we use a set of learning rates instead of a single value to train the network. This technique known as discriminative learning \citep{howard2018universal}, is used to obtain the best outcome out of different layers which might need different hyperparameters as per the feature maps in the specific layer. Discriminative learning uses the idea to train the earlier layers with a very low learning rate and the later layers with higher learning rates. This helps to avoid altering the earlier weights and massively affect the end layer weights which are more crucial for the specific case. To update the weights in the network, we require an optimizer and here we use the method of AdamW \citep{loshchilov2018decoupled} to achieve super-convergence \citep{2017arXiv170807120S} which leads to fast training over the data sample even with large architectures such as the one used for our experiments - DenseNet. With the use of a small scale dataset, the model is prone to overfitting as the decision boundaries are based on a few specific samples; in such cases, using an efficient regularisation technique helps to avoid overfitting of the model.

\section{Experiments and results}
\label{sec:expresults}

In this section, we have provided the plots and values for the best performing model out of the multiple random initialized runs of the experiment.

\subsection{Version 1: Original dataset}
%*Add graph for find\_lr*

\begin{figure}
  \includegraphics[width=1.0\linewidth]{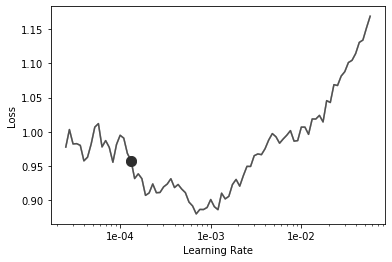}
  \caption{Graph of learning rate finder to get optimal learning rate for DenseNet-161 version 1, the black point in the graph depicts the learning rate suggested by the lr\_find method.}
  \label{fig:learningrate_1}
\end{figure}

From Figure~\ref{fig:learningrate_1}, we observe that the learning rate between the range $1e-4$ to $1e-3$ follows a non-increasing curve and can be used for optimal performance. Using this learning rate and a weight decay of $1e-3$ we trained Densenet-161 for 10 epochs. The model required 1 minute 10 seconds per epoch leading to a total time of approximately 12 minutes for training all the runs of experiments.

\begin{figure}
\centering
 \includegraphics[width=1.0\linewidth]{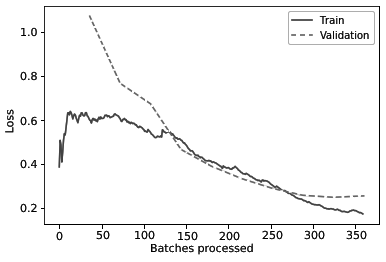}
 \caption{Plot of training and validation losses as a function of batches processed, DenseNet-161 version 1.}
\label{fig:lossVsbatch_v1}
\end{figure}

The training loss starts with a value of 0.6056 (Figure~\ref{fig:lossVsbatch_v1}) and later follows a downward path with the increase in the number of batches processed. The validation loss starts with a value above 1 (1.07) and later decreases steadily. Near the end of the batch processing, the training and the validation loss reach a value of 0.17 and 0.25 respectively.

\begin{figure}
\centering
 \includegraphics[width=1.0\linewidth]{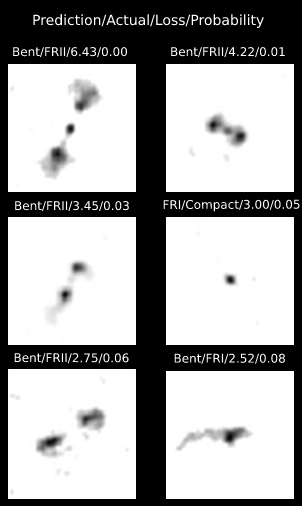}
 \caption{The top losses and incorrect classification by DenseNet-161 Version 1. Each image is $270\times270$ arcsec on a side.}
 \label{fig:TopLosses_v1}

\end{figure}

This model performed well on the validation set and the top losses can be observed in some common cases where FRII is predicted as Bent in Figure~\ref{fig:TopLosses_v1}. This figure shows the Predicted class along with the Actual (ground truth) class for the sample as well as the loss value specific to the case and the probability of the sample being predicted accurately (e.g. Bent predicted as Bent). Individual class metrics for the validation set are presented in Table~\ref{table:valid_v1}.

\begin{table}
  
    \centering
    \caption[Centre]{Validation metrics: DenseNet-161 version 1.}
    \label{table:valid_v1}
    \begin{tabular}{lcccr} % five columns, alignment for each
        \hline
        Class & Precision & Recall & F1-Score & Support\\
        \hline
        Bent & 80.2 $\pm$ 1.5 & 91.6 $\pm$ 1.9 & 86.0 $\pm$ 1.2 & 100\\
        Compact & 99.4 $\pm$ 0.6 & 99.4 $\pm$ 0.6 & 99.5 $\pm$ 0.5 & 80\\
        FRI & 94.1 $\pm$ 1.0 & 77.1 $\pm$ 1.3 & 85.2 $\pm$ 0.8 & 74\\
        FRII & 90.9 $\pm$ 1.4 & 90.5 $\pm$ 1.1 & 90.4 $\pm$ 0.9 & 143\\
        \hline
        avg/ total & 91.4 $\pm$ 0.9 & 89.7 $\pm$ 1.0 & 90.2 $\pm$ 0.9 & 397\\
        \hline
    \end{tabular}
\end{table}

The performance of the model on the test set provided an accuracy of 91.2 $\pm$ 0.6 based on multiple randomly initialized runs. The individual class metrics for the test set are presented in Table~\ref{table:Test_v1}.

\begin{table}
  
    \centering
    \caption[Centre]{Test metrics: DenseNet-161 Version 1.}
    \label{table:Test_v1}
    \begin{tabular}{lcccr} % four columns, alignment for each
        \hline
        Class & Precision & Recall & F1-Score & Support\\
        \hline
        Bent & 84.2 $\pm$ 1.3 & 94.2 $\pm$ 1.0 & 88.9 $\pm$ 0.9 & 101\\
        Compact & 97.5 $\pm$ 1.4 & 93.7 $\pm$ 1.3 & 95.7 $\pm$ 1.2 & 100\\
        FRI & 91.0 $\pm$ 0.9 & 90.0 $\pm$ 1.1 & 90.7 $\pm$ 0.7 & 100\\
        FRII & 93.5 $\pm$ 1.2 & 87.4 $\pm$ 0.8 & 90.3 $\pm$ 0.9 & 103\\
        \hline
        avg/ total & 91.7 $\pm$ 0.9 & 91.4 $\pm$ 0.9 & 91.6 $\pm$ 0.7 & 404\\
        \hline
    \end{tabular}
\end{table}

\subsection{Version 2: balanced dataset}

\begin{figure}
\centering
 \includegraphics[width=1.0\linewidth]{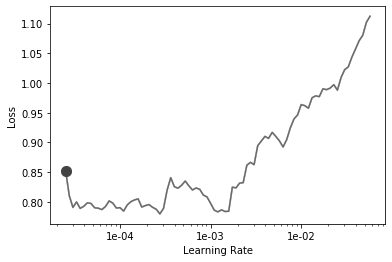}
 \caption{Graph of learning rate finder to get optimal learning rate for DenseNet-161 Version 2.}
 \label{fig:aug-lr_v2}
\end{figure}

In order to improve the performance, we tried with the balanced data sample (see section~\ref{sec:data}). Based on the learning rate finder method, here, the learning rate between the range $3e-5$ to $2e-4$ (Figure~\ref{fig:aug-lr_v2}) follows a decreasing curve and can be used for optimal performance. Using this learning rate and a weight decay of $1e-3$, we trained the Densenet-161 pre-trained model with the upsampled classes in the dataset for 10 epochs. The model took a time of 1 minute 38 seconds per epoch leading to a total time of approximately 16 minutes for training.

\begin{figure}
\centering
 \includegraphics[width=1.0\linewidth]{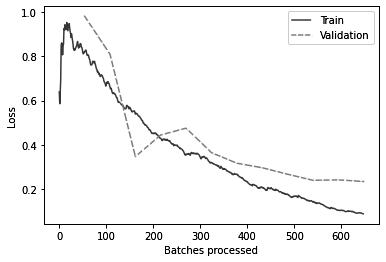}
 \caption{Plot of training and validation losses as a function of batches processed, DenseNet-161 Version 2.}
 \label{fig:lossVsbatch_v2}
\end{figure}

The training loss starts with a value of 0.5416 (Figure~\ref{fig:lossVsbatch_v2}) and later follows a steadily downward path with the increase in the number of batches processed. It reaches a minimum of 0.1588 at the end of the last epoch. The validation loss starts with a value of 0.5116, reaches a minimum value of 0.2598 near the end of the batch processing.

\begin{figure}
\centering
 \includegraphics[width=1.0\linewidth]{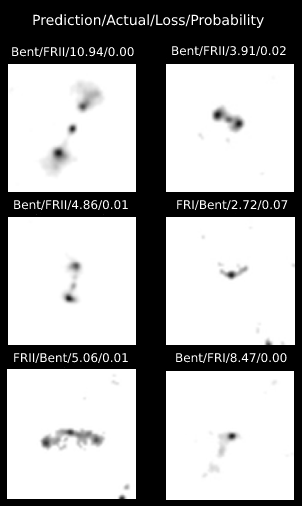}
 \caption{The top losses and incorrect classification by DenseNet-161 Version 2. Each image is $270\times270$ arcsec on a side.}
 \label{fig:aug_losses_v2}
\end{figure}

This model performed very well on the validation set with slightly better performance than Version 1 of the DenseNet-161 experiment. Individual class metrics for the validation set are presented in Table~\ref{table:Valid_v2}. The few cases where sources are classified incorrectly are shown in Figure~\ref{fig:aug_losses_v2}. Again, the figure provides values for the loss specific to the sample and the probability of correct prediction which is very low as expected with respect to the incorrect classification.

\begin{table}
  
    \centering
    \caption[Centre]{Validation metrics: DenseNet-161 Version 2.}
    \label{table:Valid_v2}
    \begin{tabular}{lcccr} % five columns, alignment for each
        \hline
        Class & Precision & Recall & F1-Score & Support\\
        \hline
        Bent & 84.0 $\pm$ 0.9 & 93.1 $\pm$ 0.8 & 88.4 $\pm$ 0.7 & 144\\
        Compact & 98.2 $\pm$ 0.5 & 98.2 $\pm$ 0.4 & 98.1 $\pm$ 0.5 & 144\\
        FRI & 92.8 $\pm$ 0.4 & 85.3 $\pm$ 0.9 & 89.1 $\pm$ 0.5 & 144\\
        FRII & 91.9 $\pm$ 1.0 & 88.8 $\pm$ 0.7 & 89.6 $\pm$ 0.8 & 144\\
        \hline
        %Mean over all classes & 92.68\% & 92.14\% & 92.31\% & 552\\
        avg/ total & 91.5 $\pm$ 0.7 & 91.1 $\pm$ 0.7 & 91.1 $\pm$ 0.6 & 576\\
        \hline
    \end{tabular}
\end{table}

For the balanced dataset, the test set provided an overall accuracy of 92.1 $\pm$ 0.4 based on multiple randomly initialized runs. The individual class metrics for the test set are presented in Table~\ref{table:Test_v2}.

\begin{table}
  
    \centering
    \caption[Centre]{Test metrics: DenseNet-161 Version 2.}
    \label{table:Test_v2}
    \begin{tabular}{lcccr} % four columns, alignment for each
        \hline
        Class & Precision & Recall & F1-Score & Support\\
        \hline
        Bent & 86.3 $\pm$ 1.2 & 87.5 $\pm$ 0.9 & 86.7 $\pm$ 1.0 & 101\\
        Compact & 98.8 $\pm$ 0.6 & 98.8 $\pm$ 0.5 & 98.7 $\pm$ 0.5 & 100\\
        FRI & 92.6 $\pm$ 0.5 & 92.7 $\pm$ 0.7 & 92.7 $\pm$ 0.6 & 100\\
        FRII & 91.3 $\pm$ 0.7 & 89.8 $\pm$ 1.3 & 90.2 $\pm$ 0.9 & 103\\
        \hline
        avg/ total & 91.9 $\pm$ 0.9 & 91.8 $\pm$ 1.2 & 91.8 $\pm$ 1.0 & 404\\
        \hline
    \end{tabular}
\end{table}

\section{Discussion}
\label{sec:discussion}

\begin{table*}
    \centering
    \caption[Centre]{Performance of various classifiers. Performance denotes Precision (P), Recall (R), F1-score (F1), Accuracy (A), mean Average Precision (mAP). Original sample size denotes the sample prior to any augmentation. Final dataset size includes the dataset size after augmentation (if performed). Morphology classes denote the radio galaxy morphology classes classified with the particular classifier. For the details on six derived classes, please refer to \citet{wu19}.}
    \label{table:rg_classifiers}
    \begin{tabular}{lcccr} % five columns, alignment for each
        \hline
        Method Name & Performance & Original sample size & Final sample size & Morphology classes \\
        \hline
        %input
        \citet{aniyan17} & P: 88\% R: 86\% F1: 0.86 & 716 & 117540 & FRI, FRII, BT \\
        % 0 & Input & 3 x 224 x 224 & - & - & - & 0 & \\
        \citet{wathela18} & P: 97\% R: 97\% F1: 0.97 & 837 & 21780 & FRI, FRI, Bent, Compact \\
        \citet{ma19}  & P: 92\% R: 84\% & 1442 & 38882 + 14245 & Compact, FRI, FRII, BT, XRG, RRG \\
        \citet{2018MNRAS.476..246L} & P: 79.6\% R: 81.6\% A: 94.8\% & 21933 & 141032 & Compact, Single extended, Two-Component \\
        & & & & extended, Multi-Component extended \\
        %& & & & extended \\
        \citet{wu19} & mAP: 84\% & 11836 & 11836 (No aug.) & Six derived classes \\
        \citet{tang19} & A: 90.1\% (FIRST), & 659 & 39,796 & FRI, FRII \\
        & A: 83.9\% (NVSS) & & & \\
        \citet{10.1093/mnras/stz1289} & P: 94.4\% & 2901 & 15936 & Unresolved, FRI, FRII \\
        \hline
    \end{tabular}
\end{table*}

An important focus of our work has been on getting accurate classification with very few training examples. Success here mainly depends on the task and its complexity. Naturally, a model learns to generalize better with a wide variety of examples rather than just a large number of examples. In our initial attempt, we used a Twin network based implementation of few-shot learning for classification of radio galaxy images. Although the results are encouraging, the accuracy achieved does not match up to the state-of-the-art classification techniques that have been presented in the recent literature. The focus of our effort then shifted to trying to use other deep learning based techniques for classification, while still working with a relatively small, albeit carefully curated sample of radio galaxies.

Deep learning algorithms generate the features from the input data at runtime (training process) and this avoids the need for feature engineering. To generate useful features, these algorithms require a large amount of data compared to traditional machine learning algorithms. The general idea is to manage the sample size and architecture coherently. In our task, since the number of classes is very less, the number of different parameters per class has not been too high, and thus even with less than 1000 images per class, we have achieved accuracy around 92.1 $\pm$ 0.4. The use of techniques like cyclical learning rate, AdamW optimizer to cope with the less amount of data have helped us to obtain good results with short training times. Though DenseNet has a large number of parameters (section \ref{parameters}), our training methods help to converge the model quickly, leading to a low cost of training.

The classification of radio galaxies is an important problem and several machine learning based techniques have been developed in recent years. However, in each case, the procedure for sample selection, the number of desired output classes and the type and configuration of machine learning architecture used are different. Thus a one-to-one comparison among these works is not possible. Our classification task closely resembles the work by \citet{wathela18} (the same 4 classes - FRI, FRII, Bent, Compact - are used for classification) in which the authors achieved an overall accuracy above 94\% while using more than 21000 samples with 400 epochs for training and 12,325,792 total learnable parameters. In this work, we obtain nearly the same performance with just over 2708 samples in total and 10 epochs for training with 2,493,348 learnable parameters. Performance of other classifiers is summarized in \ref{table:rg_classifiers}.

Our experiments are motivated by the idea to get the best possible performance with as little data as possible. The classifier presented here does not outperform all the previous attempts but we do obtain comparable performance with limited data. Convolutional Neural Networks have traditionally worked well with large amounts of data but the recent developments in the training methods have helped to get excellent outcomes even with fewer samples in the dataset.

We perform upsampling as the only form of data augmentation in our experiments. This provides additional samples during training and helps to handle the class imbalance issue. The improvement in the performance of version 2 experiments from version 1 is likely due to both these factors - availability of a larger sample as well as a balanced set. We do not increase the data sample size beyond balancing the classes, mainly to keep a small data sample.

The concatenation technique for feature maps in DenseNet helps to capture diverse characteristics rapidly and thus the architecture works well with less amount of data. This has been the primary motivation to utilize DenseNet in our experiments.

%How transfer learning worked here
Since we are using a network that has already seen a lot of images and learned to distinguish between the classes, we can teach this network to classify the new classes with a few examples. The technique has been applied mostly in the same domain tasks but the use of optimal architecture and discriminative learning has helped to obtain better results in a different domain here. \cite{wu19} have discussed the possible negative impact of transfer learning when applied for different domain transfer based on the suggestions from \cite{pan2009survey}. \cite{Rosenstein2005TransferLW} suggested using an ensemble of tasks A for a better transfer to task B mainly due to strong feature space creation. The idea of negative transfer does not seem to practically hold true, as seen from the results of our experiments.

%TL vs FSL
When comparing the performance shown by Few shot learning and transfer learning experiments, we observe a few factors leading to the difference in the results.
There is a significant difference in the architecture of Twin networks and DenseNet. DenseNet is a much deeper network with more number of parameters and residual connections helping in better flow of the gradient during training. Comparatively, twin networks are shallower and less complex in terms of the number of parameters. In few-shot learning experiments, we train the networks from scratch whereas transfer learning involves the use of pretrained networks. As shown by \cite{hendrycks2019pretraining}, pre-training helps in the robustness of the model and helps to obtain a better performance. As Few-Shot Image classification is an active area of research, we can look forward to these methods performing on par or better than the transfer learning methods in the future.

%failures, which classes most confusion and why.
Overall, we observed that the `Bent' category galaxies show the highest difficulty in correct classification. In the incorrect cases, the FRII samples are classified as Bent and vice-versa. As suggested by \cite{aniyan17} this is most likely because many Bent galaxies exhibit FRII like morphology and hence may be classified incorrectly as such.

The learning rate finder method based on cyclical learning has provided a way to adapt to the optimal learning rate parameter for all the experiments. The general idea to set the batch size has an aspect of the computational architecture linked to it and thus we set the batch size as a power of 2 i.e. 32, 64 and so on. A larger batch size ($>$ 64) can lead to memory overflow but it depends on the network architecture and the data sample. We have maintained a batch size of 32 throughout the experiments. In practice, the batch size and the number of epochs are handled interdependently. The batch size of 32 along with 10 epochs instead of a batch size of 64 and less number of epochs worked better in our experiments. Another important factor, the image size is set as $224\times224$ for our experiments based on the applications of the DenseNet architecture for ImageNet dataset. In order to utilize the weights of the DenseNet obtained for ImageNet dataset, using the ImageNet stats for our data is highly encouraged. The optimizer used in our experiments - AdamW represents an improvement over the original Adam \citep{kingma2014method} optimizer and performs better for computer vision tasks. The hyperparameter setup worked best for our case as per the hardware and GPU support but it can be adjusted, for different and possibly somewhat better results.

%Impact and contribution for the future surveys
Our system can be very useful to classify the objects of any current or future radio continuum survey. The dataset, experiments and the models are available at \href{https://github.com/kiryteo/RG_Classification_code}{https://github.com/kiryteo/RG\_Classification\_code}. Users of the system can try to get their own small and curated dataset for the training, tweak the hyperparameters mentioned above to get their own results. Since we are providing the trained model for the inference, users can try using the model as a pre-trained model, perform hyperparameter tuning and get improved results. This approach is consistent with the domain transfer idea from transfer learning. With active ongoing research for the optimization of neural network architectures, applying the techniques presented in the paper with another architecture can lead to independent results with the same training data.
\vspace{-4mm}
\smallskip

\section{Conclusions}
\label{sec:conclusion}
The main findings of this paper are as follows:
\begin{itemize}
    \item We have explored and demonstrated techniques to achieve good performance in the classification of radio galaxies into 4 classes, even when using a relatively small number of training examples.
    \item Amongst the two techniques explored in this paper, the performance of the Transfer Learning based pre-trained network (DenseNet) is superior to that achieved using Few-Shot learning based Twin Networks.
    \item Amongst the four radio galaxy classes, the classification of bent-tail galaxies proves to be the most error-prone. In most cases, bent-tail galaxies are classified as FRII galaxies and vice versa. We surmise that this happens because most bent-tail galaxies exhibit FRII morphology and are thus being confused with their (straight tail) FRII counterparts. 
\end{itemize}

\section*{Acknowledgements}
LTG and YW acknowledge the support of the Department of Atomic Energy, Government of India, under project no. 12-R\&D-TFR-5.02-0700.

\section*{Data Availability}
The data underlying this article are available in the article and in the online {github repository}\footnote{\url{https://github.com/kiryteo/RG_Classification_code}}.

% \section{Obtaining and installing the MNRAS package}
% Some \LaTeX\ distributions come with the MNRAS package by default.
% If yours does not, you can either install it using your distribution's package manager, or download it from the Comprehensive \TeX\ Archive Network\footnote{\url{http://www.ctan.org/tex-archive/macros/latex/contrib/mnras}} (CTAN).

% The files can either be installed permanently by placing them in the appropriate directory (consult the documentation for your \LaTeX\ distribution), or used temporarily by placing them in the working directory for your paper.

% Don't change these lines

\label{lastpage}
\bibliographystyle{mnras}
\bibliography{references}

\begin{thebibliography}{}
\makeatletter
\relax
\def\mn@urlcharsother{\let\do\@makeother \do\$\do\&\do\#\do\^\do\_\do\%\do\~}
\def\mn@doi{\begingroup\mn@urlcharsother \@ifnextchar [ {\mn@doi@}
  {\mn@doi@[]}}
\def\mn@doi@[#1]#2{\def\@tempa{#1}\ifx\@tempa\@empty \href
  {http://dx.doi.org/#2} {doi:#2}\else \href {http://dx.doi.org/#2} {#1}\fi
  \endgroup}
\def\mn@eprint#1#2{\mn@eprint@#1:#2::\@nil}
\def\mn@eprint@arXiv#1{\href {http://arxiv.org/abs/#1} {{\tt arXiv:#1}}}
\def\mn@eprint@dblp#1{\href {http://dblp.uni-trier.de/rec/bibtex/#1.xml}
  {dblp:#1}}
\def\mn@eprint@#1:#2:#3:#4\@nil{\def\@tempa {#1}\def\@tempb {#2}\def\@tempc
  {#3}\ifx \@tempc \@empty \let \@tempc \@tempb \let \@tempb \@tempa \fi \ifx
  \@tempb \@empty \def\@tempb {arXiv}\fi \@ifundefined
  {mn@eprint@\@tempb}{\@tempb:\@tempc}{\expandafter \expandafter \csname
  mn@eprint@\@tempb\endcsname \expandafter{\@tempc}}}

\bibitem[\protect\citeauthoryear{{Alhassan}, {Taylor}  \& {Vaccari}}{{Alhassan}
  et~al.}{2018}]{wathela18}
{Alhassan} W.,  {Taylor} A.~R.,   {Vaccari} M.,  2018, \mn@doi [\mnras]
  {10.1093/mnras/sty2038}, \href
  {https://ui.adsabs.harvard.edu/abs/2018MNRAS.480.2085A} {480, 2085}

\bibitem[\protect\citeauthoryear{{Aniyan} \& {Thorat}}{{Aniyan} \&
  {Thorat}}{2017}]{aniyan17}
{Aniyan} A.~K.,  {Thorat} K.,  2017, \mn@doi [\apjs]
  {10.3847/1538-4365/aa7333}, \href
  {https://ui.adsabs.harvard.edu/abs/2017ApJS..230...20A} {230, 20}

\bibitem[\protect\citeauthoryear{{Baldi}, {Capetti}  \& {Massaro}}{{Baldi}
  et~al.}{2018}]{baldi17}
{Baldi} R.~D.,  {Capetti} A.,   {Massaro} F.,  2018, \mn@doi [\aap]
  {10.1051/0004-6361/201731333}, \href
  {https://ui.adsabs.harvard.edu/abs/2018A&A...609A...1B} {609, A1}

\bibitem[\protect\citeauthoryear{{Banfield} et~al.,}{{Banfield}
  et~al.}{2015}]{rgz15}
{Banfield} J.~K.,  et~al., 2015, \mn@doi [\mnras] {10.1093/mnras/stv1688},
  \href {https://ui.adsabs.harvard.edu/abs/2015MNRAS.453.2326B} {453, 2326}

\bibitem[\protect\citeauthoryear{{Becker}, {White}  \& {Helfand}}{{Becker}
  et~al.}{1995}]{first95}
{Becker} R.~H.,  {White} R.~L.,   {Helfand} D.~J.,  1995, \mn@doi [\apj]
  {10.1086/176166}, \href
  {https://ui.adsabs.harvard.edu/abs/1995ApJ...450..559B} {450, 559}

\bibitem[\protect\citeauthoryear{{Blanton}, {Gregg}, {Helfand}, {Becker}  \&
  {White}}{{Blanton} et~al.}{2000}]{blanton00}
{Blanton} E.~L.,  {Gregg} M.~D.,  {Helfand} D.~J.,  {Becker} R.~H.,   {White}
  R.~L.,  2000, \mn@doi [\apj] {10.1086/308428}, \href
  {https://ui.adsabs.harvard.edu/abs/2000ApJ...531..118B} {531, 118}

\bibitem[\protect\citeauthoryear{{Blanton}, {Gregg}, {Helfand}, {Becker}  \&
  {White}}{{Blanton} et~al.}{2003}]{blanton03}
{Blanton} E.~L.,  {Gregg} M.~D.,  {Helfand} D.~J.,  {Becker} R.~H.,   {White}
  R.~L.,  2003, \mn@doi [\aj] {10.1086/368140}, \href
  {https://ui.adsabs.harvard.edu/abs/2003AJ....125.1635B} {125, 1635}

\bibitem[\protect\citeauthoryear{Blum \& Mitchell}{Blum \&
  Mitchell}{1998}]{colt98/blum}
Blum A.,  Mitchell T.,  1998, in COLT' 98: Proceedings of the eleventh annual
  conference on Computational learning theory. pp 92--100

\bibitem[\protect\citeauthoryear{{Bock}, {Large}  \& {Sadler}}{{Bock}
  et~al.}{1999}]{sumss99}
{Bock} D.~C.~J.,  {Large} M.~I.,   {Sadler} E.~M.,  1999, \mn@doi [\aj]
  {10.1086/300786}, \href
  {https://ui.adsabs.harvard.edu/abs/1999AJ....117.1578B} {117, 1578}

\bibitem[\protect\citeauthoryear{Branco, Torgo  \& Ribeiro}{Branco
  et~al.}{2015}]{branco2015survey}
Branco P.,  Torgo L.,   Ribeiro R.,  2015, A Survey of Predictive Modelling
  under Imbalanced Distributions (\mn@eprint {arXiv} {1505.01658})

\bibitem[\protect\citeauthoryear{{Burns}}{{Burns}}{1990}]{burns90}
{Burns} J.~O.,  1990, \mn@doi [\aj] {10.1086/115307}, \href
  {https://ui.adsabs.harvard.edu/abs/1990AJ.....99...14B} {99, 14}

\bibitem[\protect\citeauthoryear{{Burns}, {Ledlow}, {Loken}, {Klypin}, {Voges},
  {Bryan}, {Norman}  \& {White}}{{Burns} et~al.}{1996}]{burns96}
{Burns} J.~O.,  {Ledlow} M.~J.,  {Loken} C.,  {Klypin} A.,  {Voges} W.,
  {Bryan} G.~L.,  {Norman} M.~L.,   {White} R.~A.,  1996, \mn@doi [\apjl]
  {10.1086/310198}, \href
  {https://ui.adsabs.harvard.edu/abs/1996ApJ...467L..49B} {467, L49}

\bibitem[\protect\citeauthoryear{{Capetti}, {Massaro}  \& {Baldi}}{{Capetti}
  et~al.}{2017a}]{capetti17a}
{Capetti} A.,  {Massaro} F.,   {Baldi} R.~D.,  2017a, \mn@doi [\aap]
  {10.1051/0004-6361/201629287}, \href
  {https://ui.adsabs.harvard.edu/abs/2017A&A...598A..49C} {598, A49}

\bibitem[\protect\citeauthoryear{{Capetti}, {Massaro}  \& {Baldi}}{{Capetti}
  et~al.}{2017b}]{capetti17b}
{Capetti} A.,  {Massaro} F.,   {Baldi} R.~D.,  2017b, \mn@doi [\aap]
  {10.1051/0004-6361/201630247}, \href
  {https://ui.adsabs.harvard.edu/abs/2017A&A...601A..81C} {601, A81}

\bibitem[\protect\citeauthoryear{{Carilli}, {Furlanetto}, {Briggs}, {Jarvis},
  {Rawlings}  \& {Falcke}}{{Carilli} et~al.}{2004}]{ska04}
{Carilli} C.~L.,  {Furlanetto} S.,  {Briggs} F.,  {Jarvis} M.,  {Rawlings} S.,
   {Falcke} H.,  2004, \mn@doi [\nar] {10.1016/j.newar.2004.09.046}, \href
  {https://ui.adsabs.harvard.edu/abs/2004NewAR..48.1029C} {48, 1029}

\bibitem[\protect\citeauthoryear{Chawla, Bowyer, Hall  \& Kegelmeyer}{Chawla
  et~al.}{2002}]{chawla2002smote}
Chawla N.~V.,  Bowyer K.~W.,  Hall L.~O.,   Kegelmeyer W.~P.,  2002, \mn@doi
  [Journal of Artificial Intelligence Research] {10.1613/jair.953}, 16, 321

\bibitem[\protect\citeauthoryear{{Cheng} et~al.,}{{Cheng}
  et~al.}{2020}]{cheng20}
{Cheng} T.-Y.,  et~al., 2020, \mn@doi [\mnras] {10.1093/mnras/staa501}, \href
  {https://ui.adsabs.harvard.edu/abs/2020MNRAS.493.4209C} {493, 4209}

\bibitem[\protect\citeauthoryear{{Condon}, {Cotton}, {Greisen}, {Yin},
  {Perley}, {Taylor}  \& {Broderick}}{{Condon} et~al.}{1998}]{nvss98}
{Condon} J.~J.,  {Cotton} W.~D.,  {Greisen} E.~W.,  {Yin} Q.~F.,  {Perley}
  R.~A.,  {Taylor} G.~B.,   {Broderick} J.~J.,  1998, \mn@doi [\aj]
  {10.1086/300337}, \href
  {https://ui.adsabs.harvard.edu/abs/1998AJ....115.1693C} {115, 1693}

\bibitem[\protect\citeauthoryear{{Dehghan}, {Johnston-Hollitt}, {Franzen},
  {Norris}  \& {Miller}}{{Dehghan} et~al.}{2014}]{dehghan14}
{Dehghan} S.,  {Johnston-Hollitt} M.,  {Franzen} T.~M.~O.,  {Norris} R.~P.,
  {Miller} N.~A.,  2014, \mn@doi [\aj] {10.1088/0004-6256/148/5/75}, \href
  {https://ui.adsabs.harvard.edu/abs/2014AJ....148...75D} {148, 75}

\bibitem[\protect\citeauthoryear{Deng, Dong, Socher, Li, Li  \& Fei-Fei}{Deng
  et~al.}{2009}]{deng2009imagenet}
Deng J.,  Dong W.,  Socher R.,  Li L.-J.,  Li K.,   Fei-Fei L.,  2009, in 2009
  IEEE conference on computer vision and pattern recognition. pp 248--255

\bibitem[\protect\citeauthoryear{{Dieleman}, {Willett}  \& {Dambre}}{{Dieleman}
  et~al.}{2015}]{dieleman15}
{Dieleman} S.,  {Willett} K.~W.,   {Dambre} J.,  2015, \mn@doi [\mnras]
  {10.1093/mnras/stv632}, \href
  {https://ui.adsabs.harvard.edu/abs/2015MNRAS.450.1441D} {450, 1441}

\bibitem[\protect\citeauthoryear{{Fanaroff} \& {Riley}}{{Fanaroff} \&
  {Riley}}{1974}]{fanriley74}
{Fanaroff} B.~L.,  {Riley} J.~M.,  1974, \mn@doi [\mnras]
  {10.1093/mnras/167.1.31P}, \href
  {https://ui.adsabs.harvard.edu/abs/1974MNRAS.167P..31F} {167, 31P}

\bibitem[\protect\citeauthoryear{{Fluke} \& {Jacobs}}{{Fluke} \&
  {Jacobs}}{2020}]{Fluke2020}
{Fluke} C.~J.,  {Jacobs} C.,  2020, \mn@doi [WIREs Data Mining and Knowledge
  Discovery] {10.1002/widm.1349}, \href
  {https://ui.adsabs.harvard.edu/abs/2020WDMKD..10.1349F} {10, e1349}

\bibitem[\protect\citeauthoryear{{Gendre} \& {Wall}}{{Gendre} \&
  {Wall}}{2008}]{gendre08}
{Gendre} M.~A.,  {Wall} J.~V.,  2008, \mn@doi [\mnras]
  {10.1111/j.1365-2966.2008.13792.x}, \href
  {https://ui.adsabs.harvard.edu/abs/2008MNRAS.390..819G} {390, 819}

\bibitem[\protect\citeauthoryear{{Gendre}, {Best}  \& {Wall}}{{Gendre}
  et~al.}{2010}]{gendre10}
{Gendre} M.~A.,  {Best} P.~N.,   {Wall} J.~V.,  2010, \mn@doi [\mnras]
  {10.1111/j.1365-2966.2010.16413.x}, \href
  {https://ui.adsabs.harvard.edu/abs/2010MNRAS.404.1719G} {404, 1719}

\bibitem[\protect\citeauthoryear{Hadsell, Chopra  \& LeCun}{Hadsell
  et~al.}{2006}]{Hadsell:2006:DRL:1153171.1153654}
Hadsell R.,  Chopra S.,   LeCun Y.,  2006, in Proceedings of the 2006 IEEE
  Computer Society Conference on Computer Vision and Pattern Recognition -
  Volume 2. CVPR '06.
IEEE Computer Society, Washington, DC, USA, pp 1735--1742,
  \mn@doi{10.1109/CVPR.2006.100}, \url {https://doi.org/10.1109/CVPR.2006.100}

\bibitem[\protect\citeauthoryear{Han, Wang  \& Mao}{Han
  et~al.}{2005}]{10.1007/11538059_91}
Han H.,  Wang W.-Y.,   Mao B.-H.,  2005, in Huang D.-S.,  Zhang X.-P.,   Huang
  G.-B.,  eds, Advances in Intelligent Computing. Springer Berlin Heidelberg,
  Berlin, Heidelberg, pp 878--887

\bibitem[\protect\citeauthoryear{He \& Garcia}{He \& Garcia}{2009}]{5128907}
He H.,  Garcia E.~A.,  2009, \mn@doi [IEEE Transactions on Knowledge and Data
  Engineering] {10.1109/TKDE.2008.239}, 21, 1263

\bibitem[\protect\citeauthoryear{He, Zhang, Ren  \& Sun}{He
  et~al.}{2016}]{he2016deep}
He K.,  Zhang X.,  Ren S.,   Sun J.,  2016, in Proceedings of the IEEE
  conference on computer vision and pattern recognition. pp 770--778

\bibitem[\protect\citeauthoryear{Hebb}{Hebb}{1949}]{hebb-organization-of-behavior-1949}
Hebb D.~O.,  1949, The organization of behavior: {A} neuropsychological theory.
Wiley, New York

\bibitem[\protect\citeauthoryear{Hendrycks, Lee  \& Mazeika}{Hendrycks
  et~al.}{2019}]{hendrycks2019pretraining}
Hendrycks D.,  Lee K.,   Mazeika M.,  2019, Proceedings of the International
  Conference on Machine Learning

\bibitem[\protect\citeauthoryear{Hinton, Osindero  \& Teh}{Hinton
  et~al.}{2006}]{hinton06}
Hinton G.~E.,  Osindero S.,   Teh Y.~W.,  2006, Neural Computation, 18, 1527

\bibitem[\protect\citeauthoryear{{Hocking}, {Geach}, {Sun}  \&
  {Davey}}{{Hocking} et~al.}{2018}]{hocking18}
{Hocking} A.,  {Geach} J.~E.,  {Sun} Y.,   {Davey} N.,  2018, \mn@doi [\mnras]
  {10.1093/mnras/stx2351}, \href
  {https://ui.adsabs.harvard.edu/abs/2018MNRAS.473.1108H} {473, 1108}

\bibitem[\protect\citeauthoryear{Howard \& Gugger}{Howard \&
  Gugger}{2020}]{Howard_2020}
Howard J.,  Gugger S.,  2020, \mn@doi [Information] {10.3390/info11020108}, 11,
  108

\bibitem[\protect\citeauthoryear{Howard \& Ruder}{Howard \&
  Ruder}{2018a}]{howard2018universal}
Howard J.,  Ruder S.,  2018a, in ACL. Association for Computational
  Linguistics, \url {http://arxiv.org/abs/1801.06146}

\bibitem[\protect\citeauthoryear{Howard \& Ruder}{Howard \&
  Ruder}{2018b}]{DBLP:journals/corr/abs-1801-06146}
Howard J.,  Ruder S.,  2018b, CoRR, abs/1801.06146

\bibitem[\protect\citeauthoryear{Huang, Liu, Van Der~Maaten  \&
  Weinberger}{Huang et~al.}{2017}]{huang2017densely}
Huang G.,  Liu Z.,  Van Der~Maaten L.,   Weinberger K.~Q.,  2017, in
  Proceedings of the IEEE conference on computer vision and pattern
  recognition. pp 4700--4708

\bibitem[\protect\citeauthoryear{{Huertas-Company} et~al.,}{{Huertas-Company}
  et~al.}{2015}]{hc15}
{Huertas-Company} M.,  et~al., 2015, \mn@doi [\apjs]
  {10.1088/0067-0049/221/1/8}, \href
  {https://ui.adsabs.harvard.edu/abs/2015ApJS..221....8H} {221, 8}

\bibitem[\protect\citeauthoryear{Ioffe \& Szegedy}{Ioffe \&
  Szegedy}{2015a}]{pmlr-v37-ioffe15}
Ioffe S.,  Szegedy C.,  2015a, in Bach F.,  Blei D.,  eds,  Proceedings of
  Machine Learning Research Vol. 37, Proceedings of the 32nd International
  Conference on Machine Learning. PMLR, Lille, France, pp 448--456, \url
  {http://proceedings.mlr.press/v37/ioffe15.html}

\bibitem[\protect\citeauthoryear{Ioffe \& Szegedy}{Ioffe \&
  Szegedy}{2015b}]{DBLP:journals/corr/IoffeS15}
Ioffe S.,  Szegedy C.,  2015b, CoRR, abs/1502.03167

\bibitem[\protect\citeauthoryear{{Johnston} et~al.,}{{Johnston}
  et~al.}{2008}]{askap08}
{Johnston} S.,  et~al., 2008, \mn@doi [Experimental Astronomy]
  {10.1007/s10686-008-9124-7}, \href
  {https://ui.adsabs.harvard.edu/abs/2008ExA....22..151J} {22, 151}

\bibitem[\protect\citeauthoryear{{Jonas} \& {MeerKAT Team}}{{Jonas} \& {MeerKAT
  Team}}{2016}]{2016mks..confE...1J}
{Jonas} J.,  {MeerKAT Team} 2016, in MeerKAT Science: On the Pathway to the
  SKA. p.~1

\bibitem[\protect\citeauthoryear{Kingma \& Ba}{Kingma \&
  Ba}{2014}]{kingma2014method}
Kingma D.~P.,  Ba J.,  2014, Adam: A Method for Stochastic Optimization, \url
  {http://arxiv.org/abs/1412.6980}

\bibitem[\protect\citeauthoryear{Koch}{Koch}{2015}]{koch2015siamese}
Koch G.,  2015.

\bibitem[\protect\citeauthoryear{{Kolesnikov}, {Beyer}, {Zhai}, {Puigcerver},
  {Yung}, {Gelly}  \& {Houlsby}}{{Kolesnikov}
  et~al.}{2019}]{2019arXiv191211370K}
{Kolesnikov} A.,  {Beyer} L.,  {Zhai} X.,  {Puigcerver} J.,  {Yung} J.,
  {Gelly} S.,   {Houlsby} N.,  2019, arXiv e-prints, \href
  {https://ui.adsabs.harvard.edu/abs/2019arXiv191211370K} {p. arXiv:1912.11370}

\bibitem[\protect\citeauthoryear{Krawczyk}{Krawczyk}{2016}]{Krawczyk2016LearningFI}
Krawczyk B.,  2016, Progress in Artificial Intelligence, 5, 221

\bibitem[\protect\citeauthoryear{Krizhevsky, Sutskever  \& Hinton}{Krizhevsky
  et~al.}{2012}]{NIPS2012_4824}
Krizhevsky A.,  Sutskever I.,   Hinton G.~E.,  2012, in Pereira F.,  Burges C.
  J.~C.,  Bottou L.,   Weinberger K.~Q.,  eds, , Advances in Neural Information
  Processing Systems 25.
Curran Associates, Inc., pp 1097--1105

\bibitem[\protect\citeauthoryear{Lake, Salakhutdinov, Gross  \& Tenenbaum}{Lake
  et~al.}{2011}]{Lake_oneshot}
Lake B.~M.,  Salakhutdinov R.,  Gross J.,   Tenenbaum J.~B.,  2011, One shot
  learning of simple visual concepts

\bibitem[\protect\citeauthoryear{LeCun, Cortes  \& Burges}{LeCun
  et~al.}{2010}]{lecun2010mnist}
LeCun Y.,  Cortes C.,   Burges C.,  2010, ATT Labs [Online]. Available:
  http://yann.lecun.com/exdb/mnist, 2

\bibitem[\protect\citeauthoryear{LeCun, Bengio  \& Hinton}{LeCun
  et~al.}{2015}]{lecun2015deeplearning}
LeCun Y.,  Bengio Y.,   Hinton G.,  2015, \mn@doi [Nature]
  {10.1038/nature14539}, 521, 436

\bibitem[\protect\citeauthoryear{Loshchilov \& Hutter}{Loshchilov \&
  Hutter}{2019}]{loshchilov2018decoupled}
Loshchilov I.,  Hutter F.,  2019, in International Conference on Learning
  Representations. \url {https://openreview.net/forum?id=Bkg6RiCqY7}

\bibitem[\protect\citeauthoryear{{Lukic}, {Br{\"u}ggen}, {Banfield}, {Wong},
  {Rudnick}, {Norris}  \& {Simmons}}{{Lukic}
  et~al.}{2018}]{2018MNRAS.476..246L}
{Lukic} V.,  {Br{\"u}ggen} M.,  {Banfield} J.~K.,  {Wong} O.~I.,  {Rudnick} L.,
   {Norris} R.~P.,   {Simmons} B.,  2018, \mn@doi [\mnras]
  {10.1093/mnras/sty163}, \href
  {https://ui.adsabs.harvard.edu/abs/2018MNRAS.476..246L} {476, 246}

\bibitem[\protect\citeauthoryear{Lukic, Brüggen, Mingo, Croston, Kasieczka  \&
  Best}{Lukic et~al.}{2019}]{10.1093/mnras/stz1289}
Lukic V.,  Brüggen M.,  Mingo B.,  Croston J.~H.,  Kasieczka G.,   Best P.~N.,
   2019, \mn@doi [Monthly Notices of the Royal Astronomical Society]
  {10.1093/mnras/stz1289}, 487, 1729

\bibitem[\protect\citeauthoryear{{Luo}, {Li}, {Wang}, {He}, {Li}  \&
  {Zhou}}{{Luo} et~al.}{2018}]{8599448}
{Luo} C.,  {Li} X.,  {Wang} L.,  {He} J.,  {Li} D.,   {Zhou} J.,  2018, in 2018
  5th International Conference on Systems and Informatics (ICSAI). pp 361--366

\bibitem[\protect\citeauthoryear{{Ma} et~al.,}{{Ma} et~al.}{2019}]{ma19}
{Ma} Z.,  et~al., 2019, \mn@doi [\apjs] {10.3847/1538-4365/aaf9a2}, \href
  {https://ui.adsabs.harvard.edu/abs/2019ApJS..240...34M} {240, 34}

\bibitem[\protect\citeauthoryear{Maciejewski \& Stefanowski}{Maciejewski \&
  Stefanowski}{2011}]{5949434}
Maciejewski T.,  Stefanowski J.,  2011, in 2011 IEEE Symposium on Computational
  Intelligence and Data Mining (CIDM). pp 104--111,
  \mn@doi{10.1109/CIDM.2011.5949434}

\bibitem[\protect\citeauthoryear{{Mao}, {Johnston-Hollitt}, {Stevens}  \&
  {Wotherspoon}}{{Mao} et~al.}{2009}]{mao09}
{Mao} M.~Y.,  {Johnston-Hollitt} M.,  {Stevens} J.~B.,   {Wotherspoon} S.~J.,
  2009, \mn@doi [\mnras] {10.1111/j.1365-2966.2008.14141.x}, \href
  {https://ui.adsabs.harvard.edu/abs/2009MNRAS.392.1070M} {392, 1070}

\bibitem[\protect\citeauthoryear{{Mao}, {Sharp}, {Saikia}, {Norris},
  {Johnston-Hollitt}, {Middelberg}  \& {Lovell}}{{Mao} et~al.}{2010}]{mao10}
{Mao} M.~Y.,  {Sharp} R.,  {Saikia} D.~J.,  {Norris} R.~P.,  {Johnston-Hollitt}
  M.,  {Middelberg} E.,   {Lovell} J. E.~J.,  2010, \mn@doi [\mnras]
  {10.1111/j.1365-2966.2010.16853.x}, \href
  {https://ui.adsabs.harvard.edu/abs/2010MNRAS.406.2578M} {406, 2578}

\bibitem[\protect\citeauthoryear{{Martin}, {Kaviraj}, {Hocking}, {Read}  \&
  {Geach}}{{Martin} et~al.}{2020}]{martin20}
{Martin} G.,  {Kaviraj} S.,  {Hocking} A.,  {Read} S.~C.,   {Geach} J.~E.,
  2020, \mn@doi [\mnras] {10.1093/mnras/stz3006}, \href
  {https://ui.adsabs.harvard.edu/abs/2020MNRAS.491.1408M} {491, 1408}

\bibitem[\protect\citeauthoryear{Mcculloch \& Pitts}{Mcculloch \&
  Pitts}{1943}]{mcculloch43a}
Mcculloch W.,  Pitts W.,  1943, Bulletin of Mathematical Biophysics, 5, 127

\bibitem[\protect\citeauthoryear{{Miley}, {Perola}, {van der Kruit}  \& {van
  der Laan}}{{Miley} et~al.}{1972}]{miley72}
{Miley} G.~K.,  {Perola} G.~C.,  {van der Kruit} P.~C.,   {van der Laan} H.,
  1972, \mn@doi [\nat] {10.1038/237269a0}, \href
  {https://ui.adsabs.harvard.edu/abs/1972Natur.237..269M} {237, 269}

\bibitem[\protect\citeauthoryear{Nair \& Hinton}{Nair \&
  Hinton}{2010}]{nair2010rectified}
Nair V.,  Hinton G.~E.,  2010, in Proceedings of the 27th international
  conference on machine learning (ICML-10). pp 807--814

\bibitem[\protect\citeauthoryear{{Norris} et~al.,}{{Norris}
  et~al.}{2011}]{norris11}
{Norris} R.~P.,  et~al., 2011, \mn@doi [\pasa] {10.1071/AS11021}, \href
  {https://ui.adsabs.harvard.edu/abs/2011PASA...28..215N} {28, 215}

\bibitem[\protect\citeauthoryear{{Ostrovski} et~al.,}{{Ostrovski}
  et~al.}{2017}]{ostrovski17}
{Ostrovski} F.,  et~al., 2017, \mn@doi [\mnras] {10.1093/mnras/stw2958}, \href
  {https://ui.adsabs.harvard.edu/abs/2017MNRAS.465.4325O} {465, 4325}

\bibitem[\protect\citeauthoryear{Pan \& Yang}{Pan \&
  Yang}{2009}]{pan2009survey}
Pan S.~J.,  Yang Q.,  2009, IEEE Transactions on knowledge and data
  engineering, 22, 1345

\bibitem[\protect\citeauthoryear{Paszke et~al.,}{Paszke
  et~al.}{2019}]{paszke2019pytorch}
Paszke A.,  et~al., 2019, in Advances in neural information processing systems.
  pp 8026--8037

\bibitem[\protect\citeauthoryear{{Philip}, {Wadadekar}, {Kembhavi}  \&
  {Joseph}}{{Philip} et~al.}{2002}]{philip02}
{Philip} N.~S.,  {Wadadekar} Y.,  {Kembhavi} A.,   {Joseph} K.~B.,  2002,
  \mn@doi [\aap] {10.1051/0004-6361:20020219}, \href
  {https://ui.adsabs.harvard.edu/abs/2002A&A...385.1119P} {385, 1119}

\bibitem[\protect\citeauthoryear{Pratt}{Pratt}{1993}]{pratt1993discriminability}
Pratt L.~Y.,  1993, in Advances in neural information processing systems. pp
  204--211

\bibitem[\protect\citeauthoryear{{Proctor}, {de Oliveira}, {Dupke}, {de
  Oliveira}, {Cypriano}, {Miller}  \& {Rykoff}}{{Proctor}
  et~al.}{2011}]{proctor11}
{Proctor} R.~N.,  {de Oliveira} C.~M.,  {Dupke} R.,  {de Oliveira} R.~L.,
  {Cypriano} E.~S.,  {Miller} E.~D.,   {Rykoff} E.,  2011, \mn@doi [\mnras]
  {10.1111/j.1365-2966.2011.19625.x}, \href
  {https://ui.adsabs.harvard.edu/abs/2011MNRAS.418.2054P} {418, 2054}

\bibitem[\protect\citeauthoryear{{Roettiger}, {Burns}  \& {Loken}}{{Roettiger}
  et~al.}{1996}]{roettiger96}
{Roettiger} K.,  {Burns} J.~O.,   {Loken} C.,  1996, \mn@doi [\apj]
  {10.1086/178179}, \href
  {https://ui.adsabs.harvard.edu/abs/1996ApJ...473..651R} {473, 651}

\bibitem[\protect\citeauthoryear{Roh, Heo  \& Whang}{Roh
  et~al.}{2018}]{roh2018survey}
Roh Y.,  Heo G.,   Whang S.~E.,  2018, A Survey on Data Collection for Machine
  Learning: a Big Data -- AI Integration Perspective (\mn@eprint {arXiv}
  {1811.03402})

\bibitem[\protect\citeauthoryear{Rosenblatt}{Rosenblatt}{1958}]{rosenblatt1958perceptron}
Rosenblatt F.,  1958, \mn@doi [Psychological Review] {10.1037/h0042519}, 65,
  386

\bibitem[\protect\citeauthoryear{Rosenstein, Marx, Dietterich  \&
  Kaelbling}{Rosenstein et~al.}{2005}]{Rosenstein2005TransferLW}
Rosenstein M.,  Marx Z.,  Dietterich T.~G.,   Kaelbling L.~P.,  2005, in NIPS
  2005.

\bibitem[\protect\citeauthoryear{{Rudnick} \& {Owen}}{{Rudnick} \&
  {Owen}}{1976}]{rudwen76}
{Rudnick} L.,  {Owen} F.~N.,  1976, \mn@doi [\apjl] {10.1086/182030}, \href
  {https://ui.adsabs.harvard.edu/abs/1976ApJ...203L.107R} {203, L107}

\bibitem[\protect\citeauthoryear{Rumelhart, Hinton  \& Williams}{Rumelhart
  et~al.}{1986}]{Rumelhart:1986we}
Rumelhart D.~E.,  Hinton G.~E.,   Williams R.~J.,  1986, \mn@doi [Nature]
  {10.1038/323533a0}, 323, 533

\bibitem[\protect\citeauthoryear{{Schawinski}, {Zhang}, {Zhang}, {Fowler}  \&
  {Santhanam}}{{Schawinski} et~al.}{2017}]{schawinski17}
{Schawinski} K.,  {Zhang} C.,  {Zhang} H.,  {Fowler} L.,   {Santhanam} G.~K.,
  2017, \mn@doi [\mnras] {10.1093/mnrasl/slx008}, \href
  {https://ui.adsabs.harvard.edu/abs/2017MNRAS.467L.110S} {467, L110}

\bibitem[\protect\citeauthoryear{{Smith}}{{Smith}}{2015}]{2015arXiv150601186S}
{Smith} L.~N.,  2015, arXiv e-prints, \href
  {https://ui.adsabs.harvard.edu/abs/2015arXiv150601186S} {p. arXiv:1506.01186}

\bibitem[\protect\citeauthoryear{{Smith} \& {Topin}}{{Smith} \&
  {Topin}}{2017}]{2017arXiv170807120S}
{Smith} L.~N.,  {Topin} N.,  2017, arXiv e-prints, \href
  {https://ui.adsabs.harvard.edu/abs/2017arXiv170807120S} {p. arXiv:1708.07120}

\bibitem[\protect\citeauthoryear{Soekhoe, van~der Putten  \& Plaat}{Soekhoe
  et~al.}{2016}]{10.1007/978-3-319-46349-0_5}
Soekhoe D.,  van~der Putten P.,   Plaat A.,  2016, in Bostr{\"o}m H.,  Knobbe
  A.,  Soares C.,   Papapetrou P.,  eds, Advances in Intelligent Data Analysis
  XV. Springer International Publishing, Cham, pp 50--60

\bibitem[\protect\citeauthoryear{{Tan}, {Sun}, {Kong}, {Zhang}, {Yang}  \&
  {Liu}}{{Tan} et~al.}{2018}]{2018arXiv180801974T}
{Tan} C.,  {Sun} F.,  {Kong} T.,  {Zhang} W.,  {Yang} C.,   {Liu} C.,  2018,
  arXiv e-prints, \href {https://ui.adsabs.harvard.edu/abs/2018arXiv180801974T}
  {p. arXiv:1808.01974}

\bibitem[\protect\citeauthoryear{{Tang}, {Scaife}  \& {Leahy}}{{Tang}
  et~al.}{2019}]{tang19}
{Tang} H.,  {Scaife} A.~M.~M.,   {Leahy} J.~P.,  2019, \mn@doi [\mnras]
  {10.1093/mnras/stz1883}, \href
  {https://ui.adsabs.harvard.edu/abs/2019MNRAS.488.3358T} {488, 3358}

\bibitem[\protect\citeauthoryear{Wang, Raj  \& Xing}{Wang
  et~al.}{2017}]{DBLP:journals/corr/WangRX17}
Wang H.,  Raj B.,   Xing E.~P.,  2017, CoRR, abs/1702.07800

\bibitem[\protect\citeauthoryear{{Wing} \& {Blanton}}{{Wing} \&
  {Blanton}}{2011}]{wington11}
{Wing} J.~D.,  {Blanton} E.~L.,  2011, \mn@doi [\aj]
  {10.1088/0004-6256/141/3/88}, \href
  {https://ui.adsabs.harvard.edu/abs/2011AJ....141...88W} {141, 88}

\bibitem[\protect\citeauthoryear{{Wu} et~al.,}{{Wu} et~al.}{2019}]{wu19}
{Wu} C.,  et~al., 2019, \mn@doi [\mnras] {10.1093/mnras/sty2646}, \href
  {https://ui.adsabs.harvard.edu/abs/2019MNRAS.482.1211W} {482, 1211}

\bibitem[\protect\citeauthoryear{Yalniz, J{\'{e}}gou, Chen, Paluri  \&
  Mahajan}{Yalniz et~al.}{2019}]{DBLP:journals/corr/abs-1905-00546}
Yalniz I.~Z.,  J{\'{e}}gou H.,  Chen K.,  Paluri M.,   Mahajan D.,  2019, CoRR,
  abs/1905.00546

\bibitem[\protect\citeauthoryear{Yosinski, Clune, Bengio  \& Lipson}{Yosinski
  et~al.}{2014}]{yosinski2014transferable}
Yosinski J.,  Clune J.,  Bengio Y.,   Lipson H.,  2014, in Advances in neural
  information processing systems. pp 3320--3328

\makeatother
\end{thebibliography}
\bsp    % typesetting comment
\end{document}